\documentclass[twocolumn,10pt]{extarticle}
\usepackage{subcaption}
\usepackage[top=0.75in, bottom=0.75in, left=0.65in, right=0.65in]{geometry}
\usepackage{fourier}
\usepackage{cite}
\usepackage{amsmath,amssymb,amsfonts}
\usepackage{graphicx}
\usepackage{textcomp}
\usepackage{xcolor}
\usepackage{booktabs}
\usepackage{colortbl}
\let\temp\rmdefault
\usepackage{mathpazo}
\let\rmdefault\temp

\usepackage[framed,numbered]{matlab-prettifier}
\definecolor{roseRed}{HTML}{e20047}
\definecolor{coolGray}{HTML}{474747} 
\definecolor{mocha}{HTML}{a47864}
\definecolor{sage}{HTML}{8a9a5b}
\definecolor{dustyblue}{HTML}{6e8ca0}
\definecolor{terracotta}{HTML}{c87f5b}
\definecolor{lavender}{HTML}{9d8bb0}
\definecolor{darkmocha}{HTML}{7a5a4b}
\definecolor{darksage}{HTML}{667244}
\definecolor{darkdustyblue}{HTML}{526878}
\definecolor{darkterracotta}{HTML}{9e6347}
\definecolor{darklavender}{HTML}{766688}
\definecolor{winery}{HTML}{7e212a}

\definecolor{lightmocha}{HTML}{c2a190}
\definecolor{lightmocha1}{HTML}{d1b09c}
\definecolor{lightmocha2}{HTML}{e2c4b0}
\definecolor{lightmocha3}{HTML}{f0d9c7}

\definecolor{androidBlue}{HTML}{8AB4F8}
\definecolor{androidBlueLight}{HTML}{E8F0FE}
\definecolor{androidGreen}{HTML}{81C995}
\definecolor{androidGreenLight}{HTML}{E6F4EA}

\definecolor{androidRed}{HTML}{F28B82}
\definecolor{androidRedLight}{HTML}{FADAD7}

\definecolor{androidYellow}{HTML}{FDD663}
\definecolor{androidYellowLight}{HTML}{FEF7E0}

\definecolor{androidPurple}{HTML}{D7AEFB}
\definecolor{androidPurpleLight}{HTML}{F4EAFC}

\definecolor{androidOrange}{HTML}{FCAD70}
\definecolor{androidOrangeLight}{HTML}{FEEADC}

\definecolor{androidTeal}{HTML}{78D9EC}
\definecolor{androidTealLight}{HTML}{E6F6F9}

\definecolor{androidGray}{HTML}{DADCE0}
\definecolor{androidGrayLight}{HTML}{F1F3F4}

\usepackage[numbers,sort&compress]{natbib}
\usepackage[scale=0.95]{sourcecodepro}  
\usepackage{algorithm}
\usepackage{algpseudocode}
\usepackage{etoolbox}
\AtBeginEnvironment{algorithmic}{\small}  
\usepackage[colorlinks=true,linkcolor=blue,citecolor=blue,urlcolor=blue]{hyperref}
\def\BibTeX{{\rm B\kern-.05em{\sc i\kern-.025em b}\kern-.08em
    T\kern-.1667em\lower.7ex\hbox{E}\kern-.125emX}}

\usepackage[most]{tcolorbox}
\usepackage{amsmath}
\usepackage{xcolor}   

\definecolor{brandOrange}{HTML}{F05510}  

\newtcolorbox{equationbox}{
  colframe=brandOrange,    
  arc=3pt,             
  boxrule=1pt,         
  left=5pt,            
  right=5pt,           
  top=5pt,             
  bottom=5pt,          
  boxsep=0pt,          
  enhanced,            
  enlarge top by=2pt,  
  enlarge bottom by=2pt 
}

\newenvironment{multiequation}{%
  \setlength{\abovedisplayskip}{2pt}      
  \setlength{\belowdisplayskip}{2pt}      
  \setlength{\abovedisplayshortskip}{0pt} 
  \setlength{\belowdisplayshortskip}{0pt} 
  \begin{equationbox}
  \begin{equation}
    \begin{aligned}
}{%
    \end{aligned}
  \end{equation}
  \end{equationbox}
}

\begin{document}
\title{Followerstopper Revisited: Phase-space Lagrangian Controller for Traffic Decongestion}


\author{Rahul Bhadani\\
\small Electrical and Computer Engineering\\
\small The University of Alabama in Huntsville, Huntsville, AL, USA\\
\small Email: rahul.bhadani@uah.edu}

\maketitle




\begin{abstract}
This paper revisits Followerstopper, a phase-space-based control system that had demonstrated its ability to mitigate emergent traffic jams due to stop-and-go traffic during rush hour in the mixed-autonomy setting. Followerstopper was deployed on an autonomous vehicle. The controller attenuates the emanant traffic waves by regulating its velocity according to the relative distance and velocity of the leader car. While regulating the velocity, the controller also prevents the collision of the ego vehicle with the lead vehicle within the range specified by the controller's design parameter. The controller design is based on a configurable quadratic curve on relative distance-relative velocity phase-space that allows the transition of the regulated velocity from (i) no modification of input, (ii) decelerating to match the leader's velocity (iii) braking to avoid any imminent collision. In this paper, we explore the phase-space properties of Followerstopper and provide a detailed description of a nonlinear control law that regulates the reference input to Followerstopper within the physics-informed boundaries. We also provide a new discussion on the nominal control law that regulates the reference speed to  Followerstopper to avoid unrealistic and unsafe acceleration.

\end{abstract}


\section{Introduction}

Several analyses~\cite {chang2017there, saberi2020simple, ccolak2016understanding} of available data provided by the government and private agencies such as the Bureau of Public Roads show that cities are increasingly crossing new limits on road capacity. In the interest of solving the traffic congestion problem in an urban area, researchers from all around the world proposed several models to understand traffic dynamics~\cite{helbing2001traffic, chevallier2009improving, ossen2007driver, nam1998analyzing}. Several car-following models~\cite{gipps1981behavioral, ge2004stabilization, benekohal1988carsim, tang2012new, lazar2016review} were proposed to understand how consumer cars interact in traffic. Following the rich understanding of car-following models and traffic dynamics, Yuki Sugiyama and his team~\cite{sugiyama2008traffic} demonstrated that traffic congestion emerges without an infrastructure bottleneck when road capacity exceeds the limit, resulting in traffic waves. With the advent of new vehicle technologies such as Adaptive Cruise Control (ACC), intelligent driver-assisted systems, cooperative driving, and autonomous vehicles (AVs), it is considered that intelligent vehicles will be able to alleviate the problems arising from increasingly congested urban road networks~\cite {li2005ivs}. As the share of level 5 autonomous vehicles is not expected to reach 100\% for a few years, AVs will have to operate in mixed traffic composed of vehicles at various levels of automation, ranging from fully manual driving to level 5.

\begin{figure}[htpb]
    \centering
    \includegraphics[width=1\linewidth]{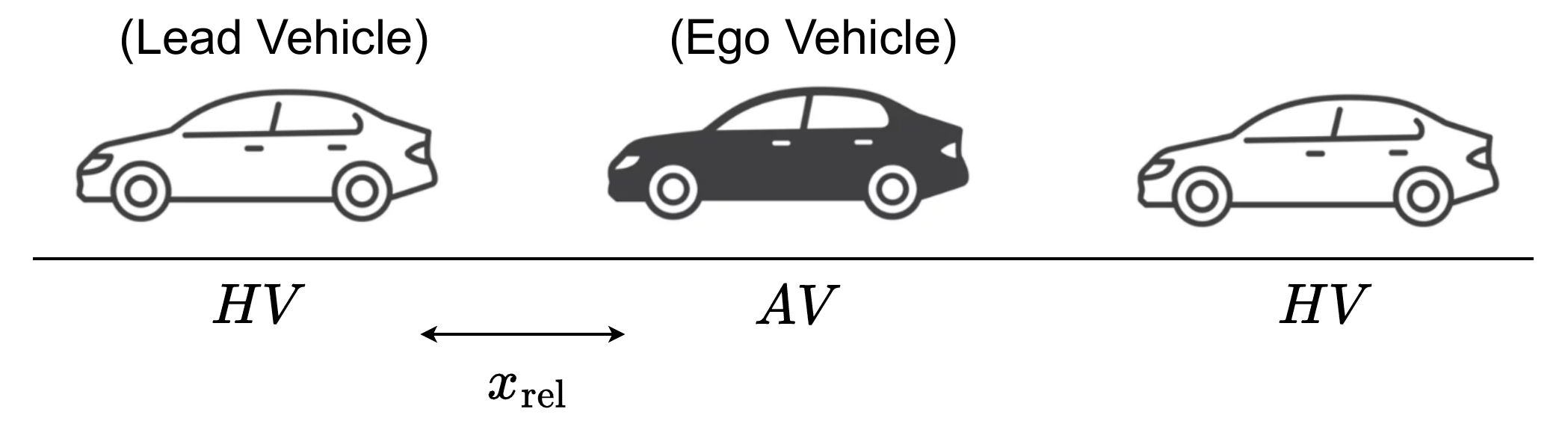}
    \caption{An autonomous vehicle following
a human-driven vehicle, which is in turn followed by another human-driven vehicle.}
    \label{fig:leader-follower-diagram}
\end{figure}

\begin{figure*}[htpb]
    \centering
    \includegraphics[width=1\linewidth]{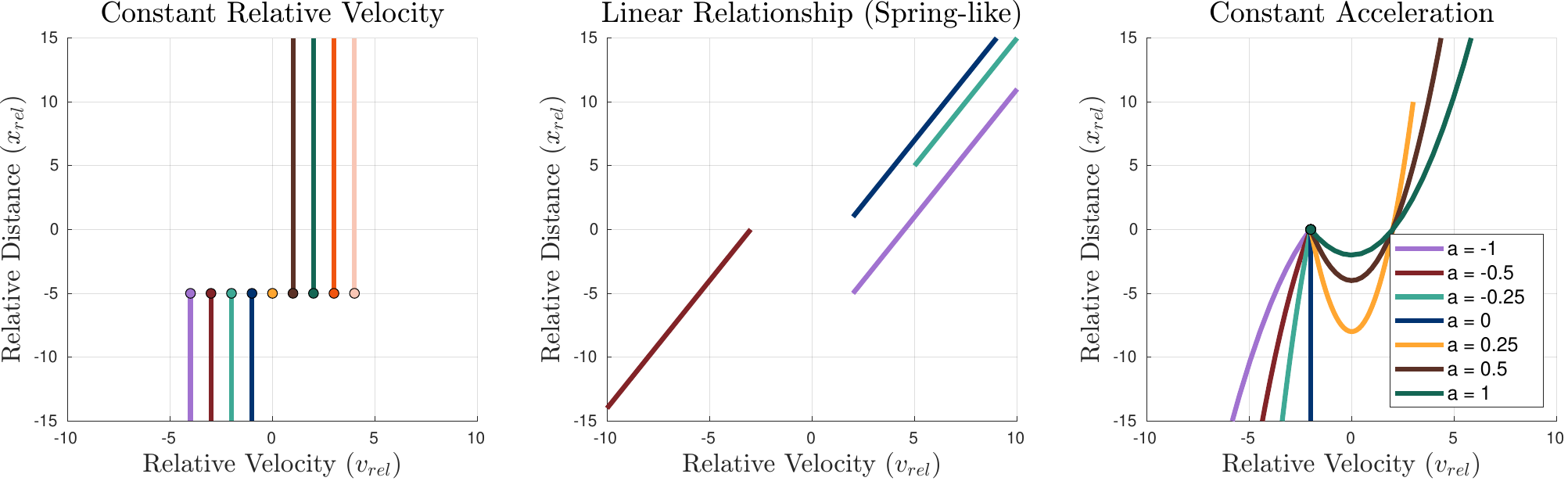}
    \caption{Phase-Space Portraits for Three Dynamical Systems Categories for Two-Car-Following Systems}
    \label{fig:phase_space_portraits}
\end{figure*}

Since the seminal Arizona Ring-road Experiment~\cite{stern2018dissipation} that demonstrated the capability of a highly automated vehicle to dissipate phantom traffic jams, the research on traffic decongestion controllers has exploded dramatically. Since the inception, methods of traffic decongestion controller have emerged such as optimal control~\cite{wang2020controllability} where authors analyize the controllability and stabilizable of a ring-road mixed traffic, model-predictive control~\cite{shang2024interaction} where authors adopt point-mass model for the dynamics of an autonomous vehicle (AV) to be controlled for traffic decongestion, and human-driven vehicle (HV) using intelligent driver
 model (IDM) with a sequence of interaction between them. In~\cite{wang2024mitigating}, authors consider the scenario of platooning with inter-vehicle communication between vehicles that is controlled by a cellular automaton model and a safe distance model. In~\cite{vinitsky2018benchmarks}, the authors presented a benchmark for testing the impact of traffic decongestion using a reinforcement learning based AV controller. In~\cite{lichtle2021fuel}, authors developed a deep reinforcement learning based AV controller to optimize fuel consumption and reduce traffic congestion simultaneously in a mixed-autonomy traffic. In another work~\cite{kreidieh2022learning}, authors used imitation learning for creating a traffic decongestion controller in a mixed-autonomy scenario. However, all of these subsequent works have been on the simulation studies and at the same time neglect practical engineering challenges such as data acquisition, real-world vehicle dynamics, and external disturbances while modeling the controller as well as the human driver.

In contrast to traditional Eulerian highway traffic controls (such as ramp metering, variable speed limits, traffic lights), traffic control via autonomous vehicle control is called Lagrangian control, where autonomous vehicles act as mobile actuators. The successful demonstration of the Lagrangian phase-space controller in Tucson, Arizona, in December 2016~\cite{stern2018dissipation}, required the design of a control strategy and significant innovation in terms of real-time data processing and filtering to handle heterogeneous in-vehicle sensor networks. However, the discussion on the prior work lacked significant discussion on the mathematical formulation of the phase-space approach that led to the design of Followerstopper. In addition, the nominal controller that regulates the reference speed of Followerstopper has not been discussed in the literature.

 \subsection{Contribution}
The contribution of this paper is towards discussing the background on phase-space-based non-linear control techniques that were used for the mitigation of traffic waves by deploying an autonomous vehicle as a Lagrangian actuator. The controller should be able to change the velocity of the autonomous vehicle smoothly from the reference velocity. The reference velocity is the velocity at which the flow dynamics of the traffic are expected to be stable. Such reference velocity comes either from an expert or from a nominal controller.  We provide further discussion of the nominal control algorithm that is currently absent in the literature. The required controller is named \textit{Followerstopper} controller, a controller that maintains a reference velocity in the flow (\textit{follower}) but can avoid the need to execute collision avoidance style braking (\textit{stopper}), since an underlying assumption of the flow dynamics is that the lead vehicle may be either speeding up or slowing down, as the AV approaches. The idea behind such a design is a controller with switching modes such that the acceleration profile does not result in further traffic wave propagation for some mode and mitigates traffic waves for some other modes. Such a design uses phase-space techniques for specifying various modes that are discussed in detail in the forthcoming section.

\section{Mixed-Autonomy Traffic Scenario}
A typical scenario in the mixed traffic where a Lagrangian control vehicle (ego vehicle) follows a human-driven vehicle, as deployed in the Arizona ring-road experiment~\cite{stern2018dissipation}, is shown in Figure~\ref{fig:leader-follower-diagram}. Typically, an ego vehicle is an autonomous vehicle (AV) or an adaptive cruise control vehicle (ACC).

To describe the experimental setup, we use the following notation for this paper:
\begin{itemize}
    \item $x_{\textrm{rel}}: [0,\infty) \rightarrow \mathbb{R}$: relative distance between the leading vehicle and the ego vehicle;
    \item $v_{\textrm{lead}}: [0,\infty) \rightarrow \mathbb{R}$: leading vehicle speed;
    \item $v_{\textrm{AV}}: [0,\infty) \rightarrow \mathbb{R}$: ego (AV) speed;
    \item $v_{\textrm{rel}}: [0,\infty) \rightarrow \mathbb{R}$: relative speed between the leading vehicle and the ego vehicle, specifically  $v_{\textrm{rel}} = v_{\textrm{lead}}-v_{\textrm{AV}}$.
    \item $v_{\textrm{cmd}}: [0,\infty) \rightarrow \mathbb{R}$: speed command for the AV;
    \item $v_{\textrm{max}}: [0,\infty) \rightarrow \mathbb{R}$: input to the nominal controller.
    \item $r:  [0,\infty) \rightarrow \mathbb{R}$: reference input to  Followerstopper.
\end{itemize}
When a subscript is not provided in the notation, we mean a generic vehicle. Mathematically, a Followerstopper controller based only on local information can be formulated as

\begin{multiequation}
\label{eq:avcontrol_form}
    v_{\text{cmd}}(t) & = f(x_{\textrm{rel}}(t), v_{\textrm{rel}}(t), v_{\textrm{lead}}(t); r) 
\end{multiequation}
Further, it should also be noted that we use synonymous terms interchangeably, such as traffic waves, shock waves, phantom traffic jams, stop-and-go waves, etc. Similarly, wave-dampening controller, traffic decongestion controller, and traffic smoothing controller refers to the controller for an AV whose goal is to smoothen out the traffic flow over a finite time period.

\subsection{Phase-space Design}
The control dynamics of  Followerstopper controller use a second-order kinematic model. For the longitudinal traffic model shown in Figure~\ref{fig:leader-follower-diagram}, the relative distance-relative velocity phase-space  ($x_{\textrm{rel}}$-$v_{\textrm{rel}}$ phase-space) falls into three distinct category:
\begin{enumerate}
    \item A constant relative velocity, i.e. $\tfrac{dv}{dt} = 0$.
    \item A linear relationship between relative distance spacing and relative velocity, i.e. $v_{\textrm{rel}} = k x_{\textrm{rel}}$
    \item A constant acceleration dynamics.
\end{enumerate}
A phase-space portrait of all three categories is shown in Figure~\ref{fig:phase_space_portraits}.

We model our autonomous car-following controller for the ego vehicle based on the constant acceleration dynamics, which provides a parabolic phase-space trajectory. We have the following kinematic equation for the dynamical system of the third category:
\begin{multiequation}
    \label{eq:v_kinematic}
v_i(t) = v_{0i}(t) + \alpha_i  t
\end{multiequation}
where $v_{0i}$ is the initial velocity, $\alpha_i$ is the acceleration of $i^{th}$ vehicle and $ t $ is the time-elapsed. The relative velocity between two vehicles can be written as 
\begin{multiequation}
\label{eq:relative_v}
v_{i-1}(t) - v_{i}(t) = v_{\textrm{rel}}(t) = v_{\textrm{rel}0} + (\alpha_{i-1} - \alpha_i)   t
\end{multiequation}
where $i-1$ indicates the index for the lead vehicle and $i$ is the index for the ego vehicle. Noting $ v_{\textrm{rel}}(t)$ as $\tfrac{dx}{dt}$, if we consider $\omega$, some specified distance as a design parameter, then we can integrate \eqref{eq:relative_v} within the limits $x=\omega$ to $x=x_{\textrm{rel}}$ from $t=0$ to $t=t$ as follows:

\begin{multiequation}
    \label{eq:integrate_relative_v}
\int_\omega^{x_{\textrm{rel}}} dx & = v_{\textrm{rel}0} \int dt + (\alpha_{i-1} - \alpha_i)\int t dt\\
x_{\textrm{rel}} - \omega & = v_{\textrm{rel}0}  t  + (\alpha_{i-1} - \alpha_i)\cfrac{t^2}{2}\\
x_{\textrm{rel}}(t)  &= \omega + v_{\textrm{rel}0}  t  + (\alpha_{i-1} - \alpha_i)\cfrac{t^2}{2}\\
\end{multiequation}

Using $t$ from \eqref{eq:relative_v}, we can write \eqref{eq:integrate_relative_v} as :

\begin{multiequation}
    \label{eq:sub_int}
x_{\textrm{rel}}(t)  &= \omega + v_{\textrm{rel}0} \cfrac{v_{\textrm{rel}}(t) - v_{\textrm{rel}0}}{(\alpha_{i-1} - \alpha_i)} + \cfrac{1}{2}\cfrac{(v_{\textrm{rel}}(t) - v_{\textrm{rel}0})^2}{(\alpha_{i-1} - \alpha_i)}
\end{multiequation}

\begin{figure}[htpb]
    \centering
    \includegraphics[width=1\linewidth]{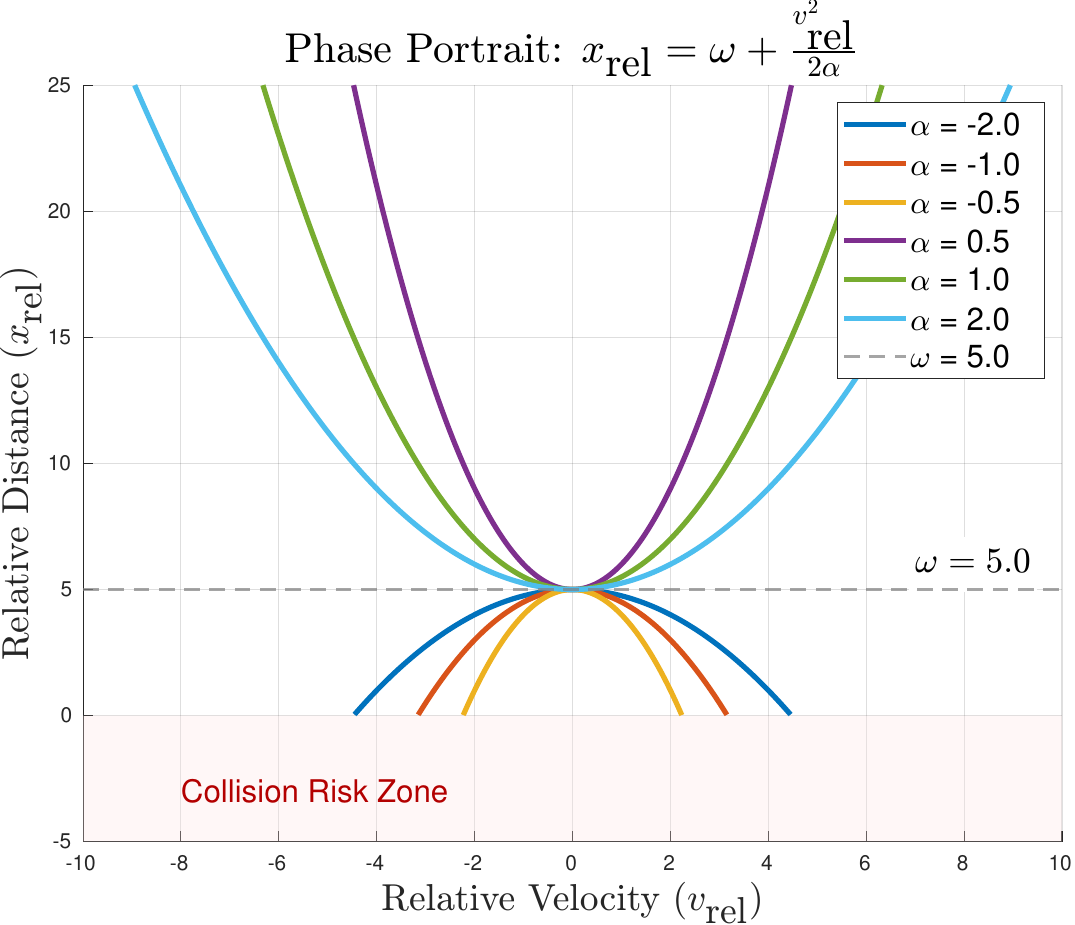}
    \caption{Phase-Space Portraits for $x_{\textrm{rel}}$-$v_{\textrm{rel}}$ with a fixed $\omega=5$ and varying $\alpha$. Vectors in the region pointing downward/leftward indicate rapid closure of the gap between the leader and the ego vehicle. Quiver plot is shown for $\alpha = 1$.}
    \label{fig:phase_portrait_sampled}
\end{figure}

If the vehicles start from rest with the same initial velocity, i.e., $v_{\textrm{rel}0} = 0$, then 
\begin{multiequation}
    \label{eq:sub_int2}
x_{\textrm{rel}}(t)  &= \omega + \cfrac{v_{\textrm{rel}}(t)^2}{2\alpha}
\end{multiequation}
with $\alpha = \alpha_{i-1} - \alpha_i$ as another design parameter. A representative phase-portrait corresponding to Equation~\eqref{eq:sub_int2} is provided in Figure~\ref{fig:phase_portrait_sampled}. Vectors denoted by arrows on the phase-portrait pointing downward/leftward indicate rapid closure of gaps. 
The gray dashed line marks the desired minimal spacing. Trajectories crossing this line indicate transitions between safe/unsafe states. Further, $\alpha > 0$

\begin{figure}[htpb]
    \centering
    \includegraphics[width=1\linewidth]{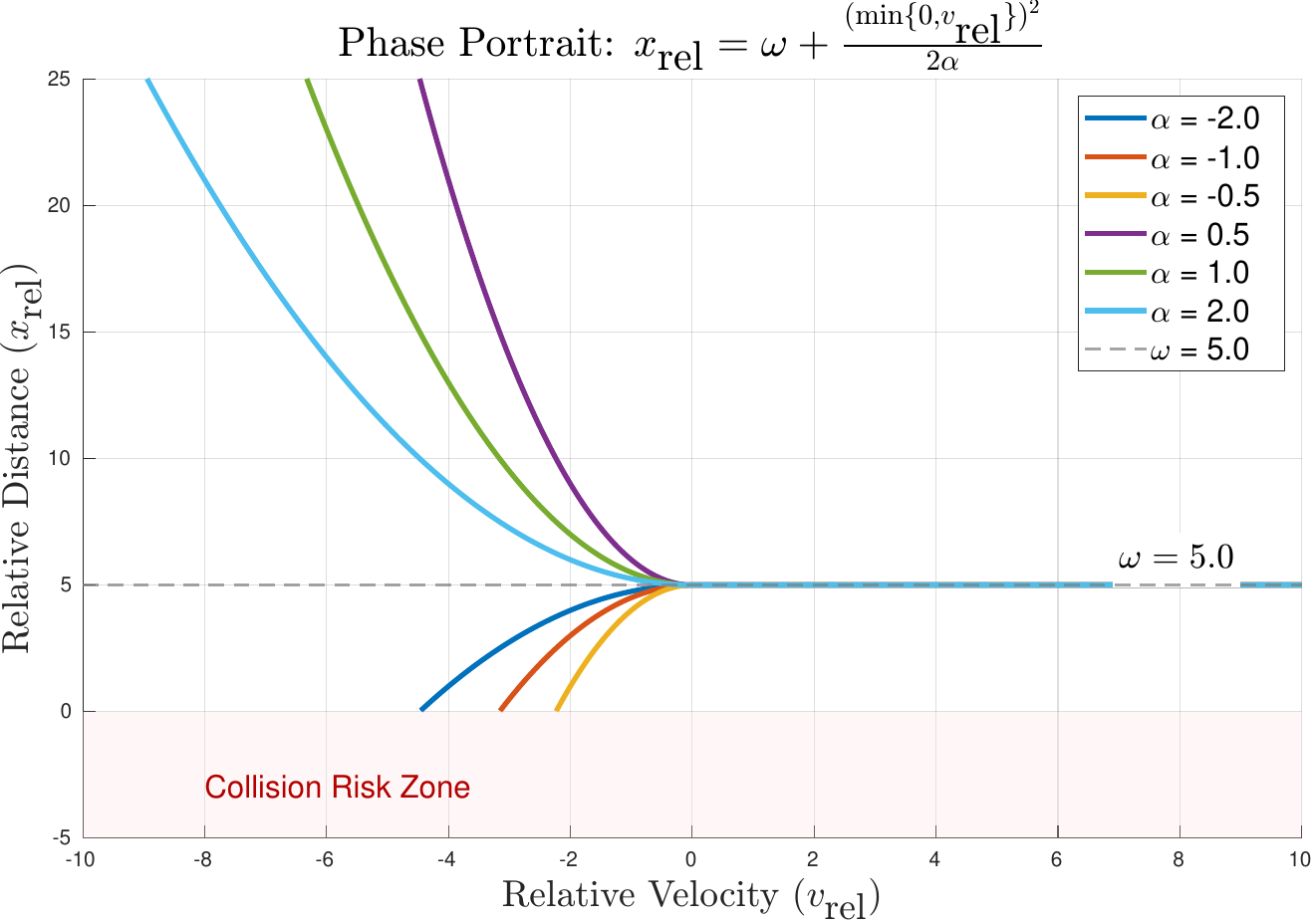}
    \caption{Phase-Space Portraits for $x_{\textrm{rel}}$-$v_{\textrm{rel}}$ with a fixed $\omega=5$ and varying $\alpha$ for Equation~\eqref{eq:sub_int3}. 
    }
    \label{fig:phase_portrait_modified}
\end{figure}

In the case of a leader-follower scenario, the negative relative velocity of the leader with respect to the ego vehicle indicates that the ego vehicle is falling behind. In such a case, the controller should command the ego vehicle with velocity $u = v_{i-1}$ to close the gap. Therefore, we modify the phase-space equation as
\begin{multiequation}
    \label{eq:sub_int3}
x_{\textrm{rel}}(t)  &= \omega + \cfrac{1}{2\alpha}(\min\{0, v_{\textrm{rel}}(t)\})^2
\end{multiequation}
A modified phase-portrait corresponding to Equation~\eqref{eq:sub_int3} is presented in Figure~\ref{fig:phase_portrait_modified}.

\section{Followerstopper: Non-linear Controller for Wave Dampening}
In order to dampen the traffic waves arising as a result of the bottleneck, the velocity controller of the form provided in Equation~\eqref{eq:avcontrol_form} needs to modulate the speed of AV to close the gap between the leader and the ego gracefully so that acceleration or deceleration doesn't amplify the traffic waves along with staying at the safe distance from the leader vehicle.
The control law is defined as follows:
\begin{multiequation}
    \label{eq:followerstopper}
& f_{\textrm{FS}}(x_{\textrm{rel}}(t), v_{\textrm{rel}}(t), v_{\textrm{lead}}(t))  = \\
& \begin{cases}
    0,  &\textrm{if} (x_{\textrm{rel}},v_{\textrm{rel}}) \in \mathcal{S}_1\\
    v(v_{\textrm{lead}}) \frac{x_{\textrm{rel}}-d_1(t)}{\bar{d}_2(t)-d_1(t)}, &\textrm{if}  (x_{\textrm{rel}},v_{\textrm{rel}}) \in \mathcal{S}_2\\
    v(v_{\textrm{lead}})+(r-v(v_{\textrm{lead}}))\frac{x_{\textrm{rel}}-d_2(t)}{d_3(t)-\bar{d}_2(t)}, &\textrm{if} (x_{\textrm{rel}},v_{\textrm{rel}}) \in \mathcal{S}_3\\
    r, &\textrm{if} (x_{\textrm{rel}},v_{\textrm{rel}}) \in \mathcal{S}_4 
    \end{cases}
\end{multiequation}
where $v : \mathbb{R} \rightarrow \mathbb{R}$ is $v(v_{\textrm{lead}}) = \min\{ \max\{v_{\textrm{lead}},0\},r\}$. Four sets $\mathcal{S}_1$, $\mathcal{S}_2$, $\mathcal{S}_3$, and $\mathcal{S}_4$ divided by three safety envelopes as are defined below:
\begin{multiequation}
     \label{eq:FS_regions}
             \mathcal{S}_1 &= \big\{(x_{\textrm{rel}},v_{\textrm{rel}}) \in \mathbb{R}^2 | 
    0 < x_{\textrm{rel}} \leq d_1(v_{\textrm{rel}}) \big\}, \\
    \mathcal{S}_2 &= \big\{(x_{\textrm{rel}},v_{\textrm{rel}}) \in \mathbb{R}^2 | d_1(v_{\textrm{rel}}) < x_{\textrm{rel}} \leq d_2(v_{\textrm{rel}}) \big\}, \\
    \mathcal{S}_3 &= \big\{(x_{\textrm{rel}},v_{\textrm{rel}}) \in \mathbb{R}^2 |d_2(v_{\textrm{rel}}) < x_{\textrm{rel}} \leq d_3(v_{\textrm{rel}}) \big\}, \\
    \mathcal{S}_4 &= \big\{(x_{\textrm{rel}},v_{\textrm{rel}}) \in \mathbb{R}^2 |d_3(v_{\textrm{rel}}) < x_{\textrm{rel}} \big\}.
\end{multiequation}

\begin{figure}[tpb]
    \centering
    \includegraphics[trim={1.4cm 0.2cm 1.4cm 0.2cm},clip, width=1.0\linewidth]{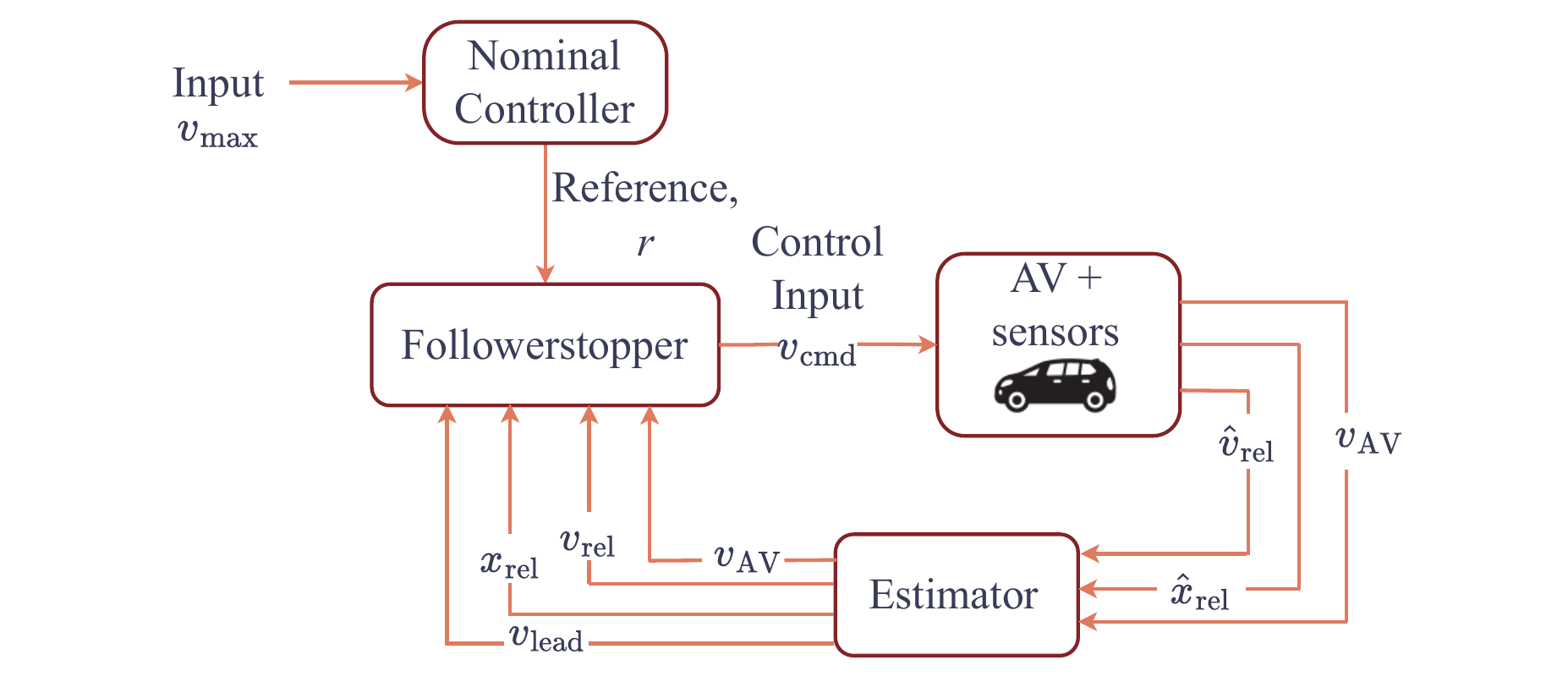}
    \caption{A schematic diagram of  Followerstopper controller and other components used for the AV control.}
    \label{fig:fs_block}
\end{figure}

The phase-portrait specified in Equation~\eqref{eq:sub_int3} separates switching region with $d_j : \mathbb{R} \rightarrow \mathbb{R}$ are:
\begin{multiequation}
    d_j(v_{\textrm{rel}}) = \omega_j + \frac{1}{2\alpha_j} \min \{0,v_{\textrm{rel}}\}^2, \quad j = 1,2,3,
    \label{eq:FS_envelopes}
\end{multiequation}
where $\omega_1 = 4.5$, $\omega_2= 5.25$, $\omega_3 = 6.0$, $\alpha_1 = 1.5$, $\alpha_2 = 1$, $\alpha_3 = 0.5$. Through our controller design, we control the velocity of the Lagrangian control vehicle (ego vehicle),
through the control input $u$ as shown in~\ref{fig:fs_block}.

\begin{figure}[tpb]
    \centering
    \includegraphics[width=1.0\linewidth]{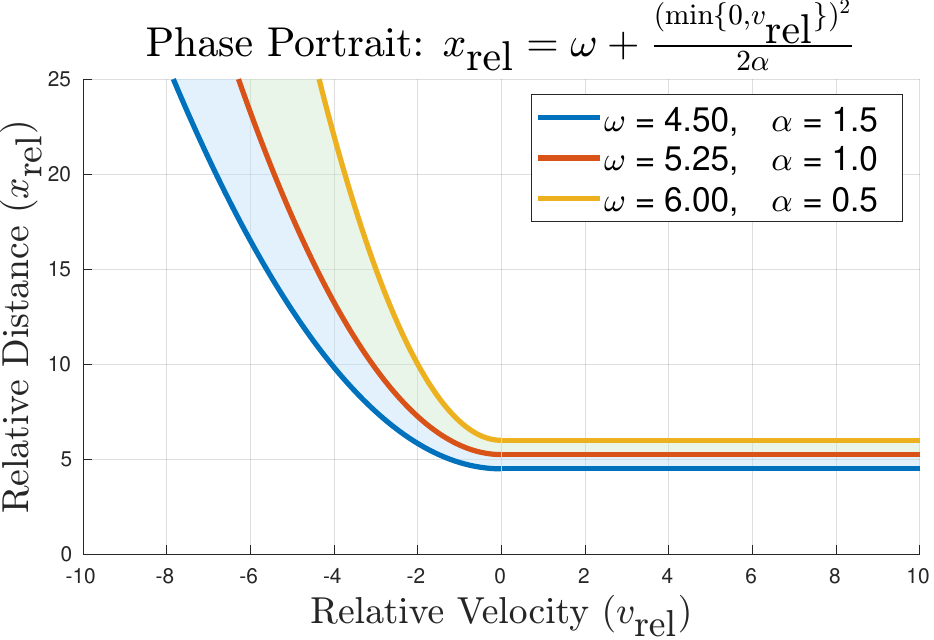}
    \caption{Switching regions allows a smoother transition from speeding up to catch up to the lead vehicle, to decelerating to match the lead vehicle’s velocity, to braking to avoid an impending collision.}
    \label{fig:switching_region}
\end{figure}

The switching region as specified by Equation~\eqref{eq:FS_envelopes} is illustrated in Figure~\ref{fig:switching_region}.

\subsection{Nominal Controller}
The overall structure of the controller, as illustrated in Figure~\ref{fig:fs_block}, shows that it requires an input speed that serves as a basis for the reference speed to the AV. However, to constrain the input within the bounds of physics so as not to cause any abrupt acceleration or deceleration (or example,  at one time-step if the input is velocity of $4m/s$ and then in the next time step it is $12.0m/$), the nominal controller further modulates the reference speed before feeding it into Followerstopper. The algorithm used for the nominal control is specified in Algorithm~\ref{alg:nominal_control}.
\begin{algorithm}[H]
\caption{Nominal Control Reference Velocity Calculation}
\label{alg:nominal_control}
\begin{algorithmic}[1]
\Require{
    $\mathit{max\_speed}$: Desired velocity (m/s), 
    $\mathit{vel}$: Current velocity (m/s),
    $\mathit{max\_accel}$: Maximum acceleration ($\textrm{m}/\textrm{s}^2$),
    $\mathit{max\_decel}$: Maximum deceleration ($\textrm{m}/\textrm{s}^2$)
}
\Ensure{$r$: Reference velocity for Followerstopper controller (m/s)}

\State $\mathit{persistent}\ y \gets 0$ \Comment{Initialize internal state variable}
\State $\mathit{dt} \gets 0.05$ \Comment{Time step (s)}

\If{$y > \mathit{max\_speed} + 1$}
    \State $y \gets \max(\mathit{max\_speed},\ y - |\mathit{max\_decel}| \cdot \mathit{dt})$ \label{line:decel}
\ElsIf{$y < \mathit{max\_speed} - 1$}
    \State $y \gets \min(\mathit{max\_speed},\ y + \mathit{max\_accel} \cdot \mathit{dt})$ \label{line:accel}
\Else
    \State $y \gets \mathit{max\_speed}$ \Comment{Maintain target speed} \label{line:maintain}
\EndIf

\If{$y < 2 \ \mathbf{and} \ \mathit{max\_speed} > 2$} \Comment{Speed floor constraints}
    \State $y \gets 2$
\ElsIf{$y < 1 \ \mathbf{and} \ \mathit{max\_speed} > 1$}
    \State $y \gets 1$
\EndIf

\State $r \gets \min(\max(y,\ \mathit{vel} - 1.0),\ \mathit{vel} + 2.0)$ \Comment{Output smoothing} \label{line:smooth}
\State \Return $r$
\end{algorithmic}
\end{algorithm}

\subsection{Implementation Details}
\begin{figure*}[htpb]
\centering
\includegraphics[trim={0.0cm 0.0cm 0.0cm 0.0cm},clip, width=1.0\linewidth]{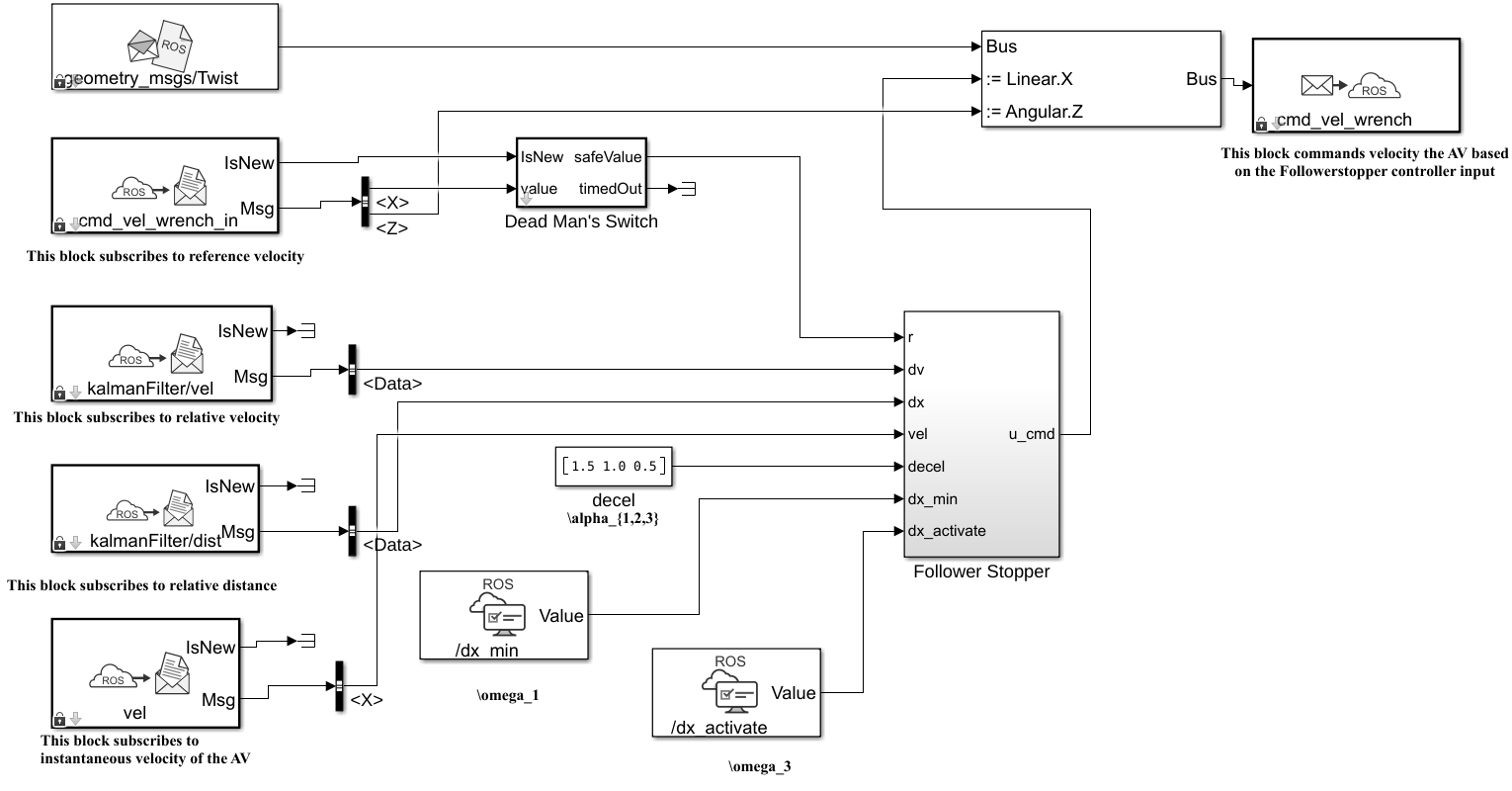}
\caption{Followerstopper Simulink Model. The function block specifies the logic from Equation~\eqref{eq:followerstopper}.}
\label{fig:FS_Simulink}
\end{figure*}

\begin{figure*}[htpb]
\centering
\includegraphics[trim={0.0cm 0.0cm 0.0cm 0.0cm},clip, width=1.0\linewidth]{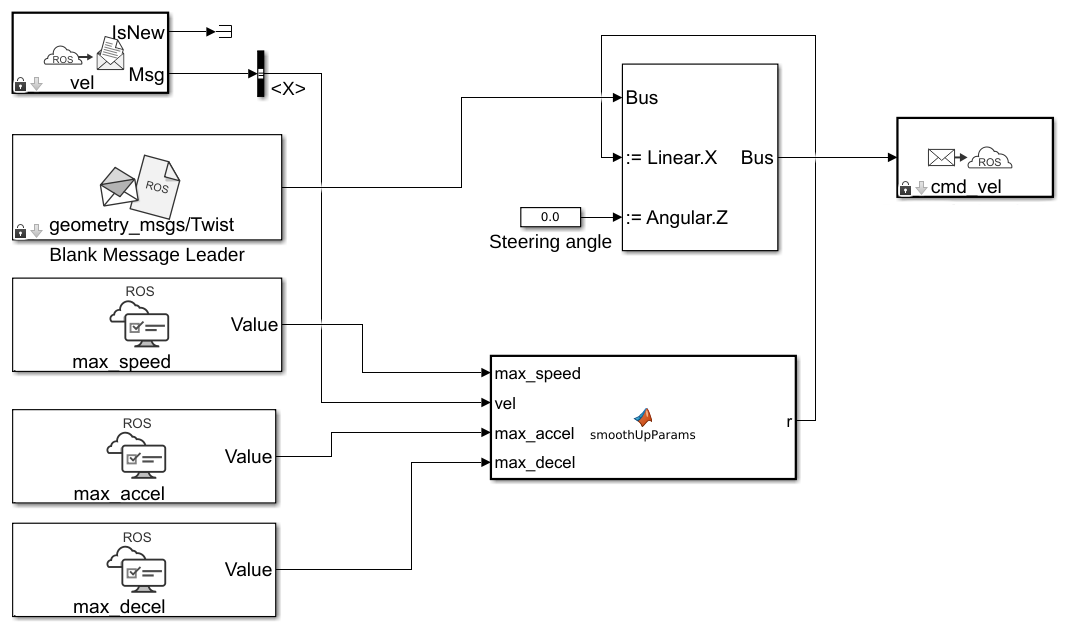}
\caption{Nominal Controller Implementation in smoothUpParams Simulink Block as specified in Algorithm~\ref{alg:nominal_control}. cmd\_vel publisher block is mapped to cmd\_vel\_wrench\_in subscriber block from Followerstopper Simulink Model shown in Figure~\ref{fig:FS_Simulink}.}
\label{fig:SmoothUpParams}
\end{figure*}

We used MATLAB's Robot Operating System (ROS) Toolbox~\cite{quigley2009ros} and Simulink to implement Followerstopper defined in Equation~\eqref{eq:followerstopper}. The ego vehicle used for our experiment was the University of Arizona's CAT Vehicle that uses ROS for high-level control and at a lower level, a ROS2JAUS interface~\cite{morley2013generating} sends a command to the vehicle's actuator. Using ROS Toolbox from Simulink, we were able to generate native ROS code to transfer to CAT Vehicle's hardware.  A snapshot of Simulink blocks to implement Followerstopper and Nominal Controller is provided in Figure~\ref{fig:FS_Simulink} and Figure~\ref{fig:SmoothUpParams}, respectively.

\section{Results}
\begin{figure*}[htpb]
\centering
\includegraphics[trim={0.0cm 0.0cm 0.0cm 0.0cm},clip, width=1.0\linewidth]{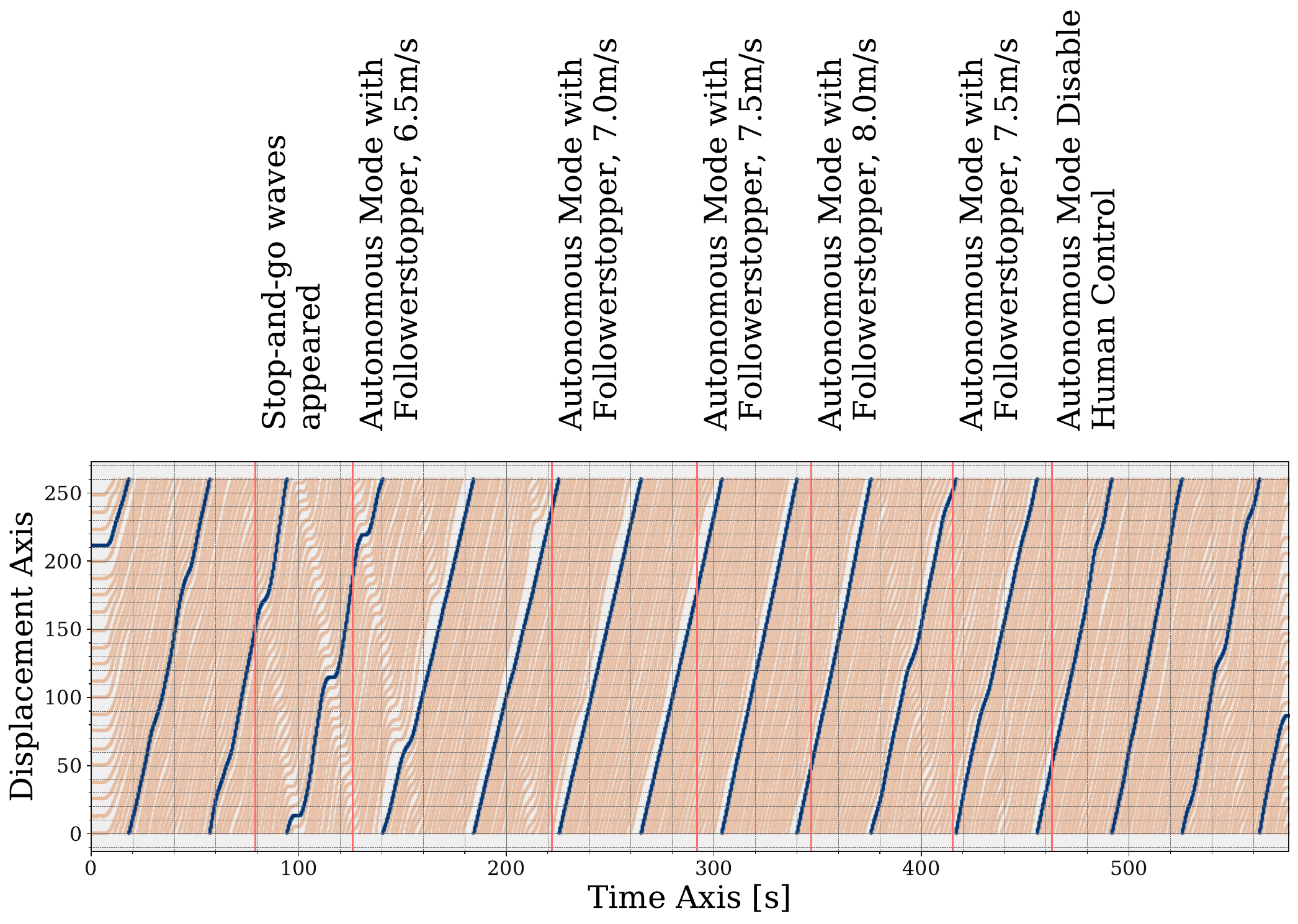}
\caption{Annotated time-space diagram for Arizona-ring road experiment with Followerstopper's wave-dampening capability.}
\label{fig:timespace_diagram}
\end{figure*}

While the overall result of the Arizona-ring Road Experiment, where Followerstopper controller was tested, is available in~\cite{stern2018dissipation}, we briefly present the time-space diagram of the experiment in Figure~\ref{fig:timespace_diagram} for the sake of completeness.
The time-space diagram shows the formation of stop-and-go waves at around $t=79$ s. At $t=126s$ from the experiment, we activated the autonomous mode of CAT Vehicle, and Followerstopper took over the control of the vehicle. At this point, user-specified input for max speed, as the input to the Nominal Controller, was $6.5m/s,$ which was provided by an expert human observing the ring traffic from a bird-eye view. It took approximately 30-35 seconds before we could observe the impact of Followerstopper's wave-dampening capability. For the next several minutes, we varied the user-specified reference velocity to assess the impact of the reference velocity setpoint on the ring traffic. A list of various reference velocity setpoints used during the experiment is shown in Table~\ref{tab:reference_velocity_table}. We observed the best performance of the wave-dampening effect at the reference speed of $7.5m/s$.

\begin{table}[h]
\centering
\normalsize
\begin{tabular}{cc}
\toprule
\rowcolor{androidBlueLight}
\textbf{\textcolor{roseRed}{Time into the Experiment}} & \textbf{\textcolor{roseRed}{Reference Velocity}} \\
\midrule
\rowcolor{androidGreenLight}
\textcolor{darksage}{0--126\,s} & \textcolor{darksage}{Manual driving} \\
\rowcolor{androidYellowLight}
\textcolor{darkmocha}{126--222\,s} & \textcolor{darkmocha}{6.5\,m/s} \\
\rowcolor{androidPurpleLight}
\textcolor{darklavender}{222--292\,s} & \textcolor{darklavender}{7.0\,m/s} \\
\rowcolor{androidOrangeLight}
\textcolor{darkterracotta}{292--347\,s} & \textcolor{darkterracotta}{7.5\,m/s} \\
\rowcolor{androidTealLight}
\textcolor{darkdustyblue}{347--415\,s} & \textcolor{darkdustyblue}{8.0\,m/s} \\
\rowcolor{androidRedLight}
\textcolor{winery}{415--463\,s} & \textcolor{winery}{7.5\,m/s} \\
\rowcolor{androidGrayLight}
\textcolor{coolGray}{$>$463\,s} & \textcolor{coolGray}{Manual driving} \\
\bottomrule
\end{tabular}
\caption{Varying reference velocity for different time intervals specified during the Arizona ring-road experiment to demonstrate the wave-dampening capability of Followerstopper}
\label{tab:reference_velocity_table}
\end{table}

\begin{figure*}
    \centering
    \includegraphics[width=1\linewidth]{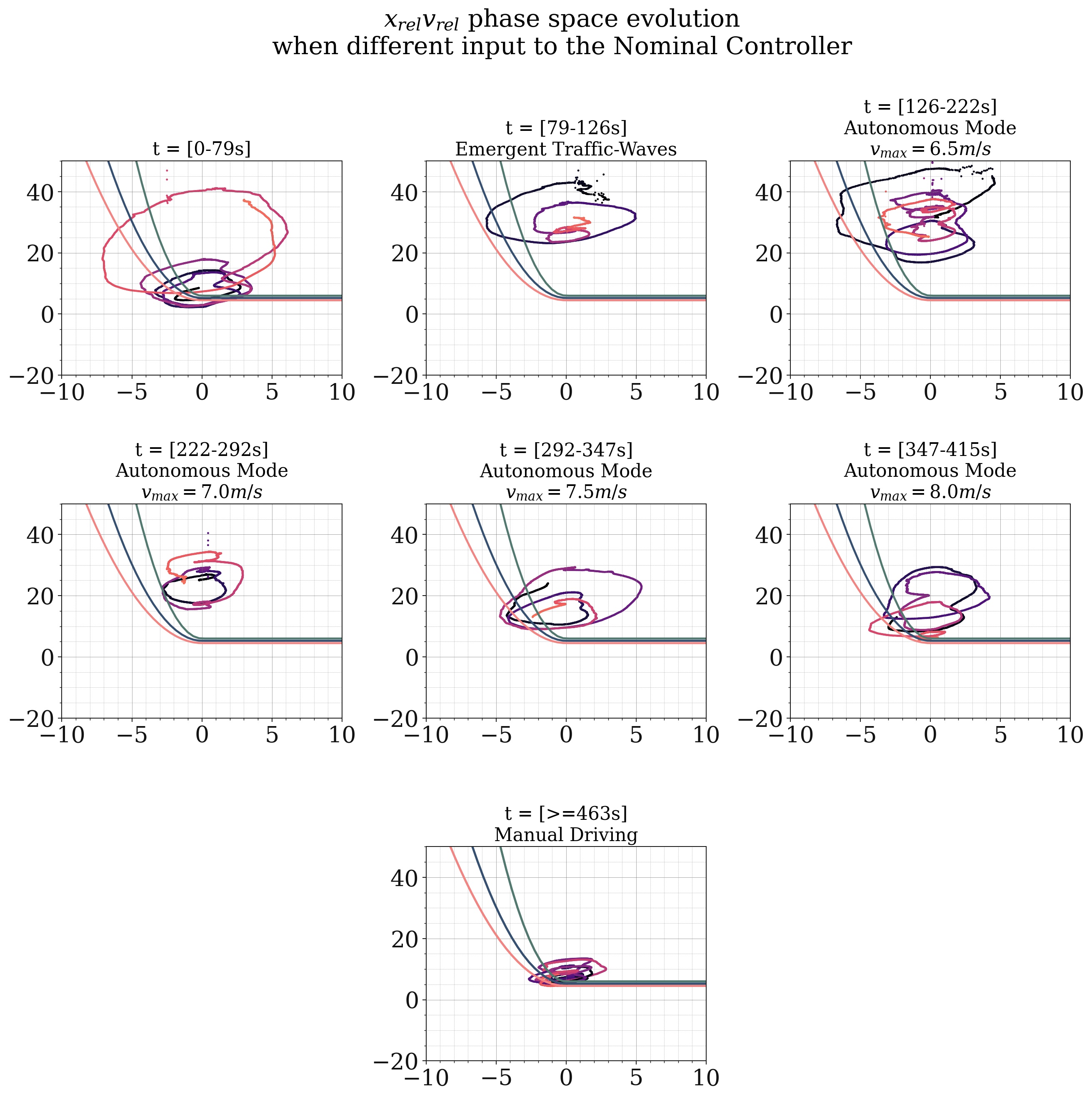}
    \caption{$x_{\textrm{rel}}-v_{\textrm{rel}}$ evolution overlaid with the switching boundary.}
    \label{fig:phase_plot_diagram}
 
\end{figure*}

We also provide $x_{\textrm{rel}}-v_{\textrm{rel}}$ phase-space diagram for the AV in Figure~\ref{fig:phase_plot_diagram}. We see that the phase-space curve is not seen in the Collision Risk Zone when the autonomous mode was active under Followerstopper control. Thus, we conclude that for the speed regime under which the autonomous vehicle was operating, its behavior was safe.

%

   \vfill

\section{Data and Code Availability}
Data generated as a part of Arizona Ring-road Experiment is available at \url{http://hdl.handle.net/1803/9358}. Code used to generate figures along with Simulink models of Followerstopper are available on \url{https://github.com/AARC-lab/Followerstopper}. \textbf{Appendix~\ref{sec:appXA}} provides m-file implementations of Followerstopper and Nominal controller algorithm (provided as \texttt{smoothUpParams} function).

\section{Conclusion and Future Work}
In this paper, we provided the mathematical derivation and complete formulation of Followerstopper controller that demonstrated its ability to dampen stop-and-go traffic waves in the Arizona Ring-road Experiment. The nominal controller that regulates the reference input to Followerstopper was formally presented for the first time, addressing a gap in the literature.

A natural extension of Followerstopper can utilize a control barrier certificate, as Equation~\eqref{eq:FS_envelopes} acts as a barrier function. The work presented in this manuscript can be extended to take into account uncertainty in state estimation to evaluate controller stability and modify its formulation.

{
\let\clearpage\relax
\small
\bibliographystyle{IEEEtran}
\bibliography{references}
}


\onecolumn
\appendix

\section{Simulink Models and Code}
\label{sec:appXA}

\subsection{m-file Implementation of Followerstopper}

\begin{tcolorbox}[
    enhanced,
    colback=androidBlueLight,
    colframe=androidBlue,
    arc=5pt,
    boxrule=1pt,
    title=\textbf{Followertopper Controller},
    fonttitle=\bfseries,
    coltitle=black,
    top=10pt,
    bottom=8pt,
    left=8pt,
    right=8pt,
    attach boxed title to top left={xshift=10pt, yshift=-\tcboxedtitleheight/2},
    boxed title style={
        colback=androidBlue,    
        colframe=androidBlue,
        arc=3pt,
        boxrule=0pt,
        left=6pt, right=6pt,
        top=3pt, bottom=3pt
    }
]
\begin{lstlisting}[
    style=Matlab-editor,
    basicstyle=\ttfamily\footnotesize,
    numbers=left,
    numberstyle=\tiny,
    xleftmargin=2em,
    framexleftmargin=1.5em
]
function u_cmd = FollowerStopper(r,dx,dv,v_AV, dx_min,dx_activate,decel)
%#codegen
% Safety controller, based on quadratic bands.
% Input: r = desired velocity (from other models)
%          dx = estimate of gap to vehicle ahead
%          dv = estimate of time derivative of dx (= velocity difference)
%         v_AV = velocity of AV
%      dx_min = minimum distance (omega_1)
% dx_activate = distance below which controller does something (omega_3)
%       decel = vector of three deceleration values for parabolas
% Out:  u_cmd = actual velocity commanded (always u_cmd<=U)
%
% Controller-specific parameters
dx_mid = (dx_min+dx_activate)/2; % mid distance, where v_lead is commanded
% Lead vehicle velocity
v_lead = v_AV+dv; % velocity of lead vehicle
v_lead = max(v_lead,0); % lead vehicle cannot go backwards
v = min(r,v_lead); % safety velocity cannot exceed desired velocity U
% Treatment of positive dv-values
dv = min(dv,0); % domains for dv>0 same as dv=0
% For given dv, dx-values of band boundaries
dx1 = dx_min+1/(2*decel(1))*dv.^2;
dx2 = dx_mid+1/(2*decel(2))*dv.^2;
dx3 = dx_activate+1/(2*decel(3))*dv.^2;
% Actual commanded velocity
u_cmd = double((dx1<dx&dx<=dx2).*(v.*(dx-dx1)./(dx2-dx1))+...
    (dx2<dx&dx<=dx3).*(v+(r-v).*(dx-dx2)./(dx3-dx2))+...
    (dx3<dx).*r);

if( dx > 16.0 )
    u_cmd = r;
end
\end{lstlisting}
\end{tcolorbox}
\vfill

\subsection{m-file Implementation of Nominal Controller}

\begin{tcolorbox}[
    enhanced,
    colback=androidBlueLight,
    colframe=androidBlue,
    arc=5pt,
    boxrule=1pt,
    title=\textbf{Nominal Controller},
    fonttitle=\bfseries,
    coltitle=black,
    top=10pt,
    bottom=8pt,
    left=8pt,
    right=8pt,
    attach boxed title to top left={xshift=10pt, yshift=-\tcboxedtitleheight/2},
    boxed title style={
        colback=androidBlue,    
        colframe=androidBlue,
        arc=3pt,
        boxrule=0pt,
        left=6pt, right=6pt,
        top=3pt, bottom=3pt
    }
]
\begin{lstlisting}[
    style=Matlab-editor,
    basicstyle=\ttfamily\footnotesize,
    numbers=left,
    numberstyle=\tiny,
    xleftmargin=2em,
    framexleftmargin=1.5em
]
function r = smoothUpParams(max_speed,vel,max_accel,max_decel)
%#codegen
% Input: max_speed = desired velocity from user
%          vel = current velocity of the vehicle
%          max_accel = maximum permissible acceleration
%         max_decel = maximum permissible deceleration
%      dx_min = minimum distance (omega_1)
% dx_activate = distance below which controller does something (omega_3)
%       decel = vector of three deceleration values for parabolas
% Out:  r =reference velocity to the Followerstopper

persistent y

if isempty(y)
    y=0;
end

dt=0.05;
if( y > max_speed + 1 )
    y = max(max_speed,y - abs(max_decel)*dt);
elseif( y < max_speed - 1)
    y = min(max_speed,y + max_accel*dt);
else
    y = double(max_speed);
end

if (y < 2 && max_speed > 2)
    y = 2;
elseif(y < 1 && max_speed > 1)
    y = 1;
end

r =min(max(y,vel-1.0),vel+2.0);
\end{lstlisting}
\end{tcolorbox}


\end{document}